\documentclass[%
 reprint,
superscriptaddress,
 amsmath,amssymb,
 aps,
prb,
]{revtex4-1}

\usepackage{graphicx}
\usepackage{dcolumn}
\usepackage{bm}
\usepackage{amsmath,amssymb}
\usepackage{graphicx}
\usepackage{color}

\usepackage{hyperref}

\begin{document}

\preprint{AIP/123-QED}

\title{Direct laser printing of high-resolution physically unclonable anti-counterfeit labels}

\author{V.~Lapidas}
\affiliation{Institute of Automation and Control Processes, Far Eastern Branch, Russian Academy of Science, Vladivostok 690041, Russia}
\affiliation{Far Eastern Federal University, 8 Sukhanova Str., 690091 Vladivostok, Russia}

\author{A.~Zhizhchenko}
\affiliation{Institute of Automation and Control Processes, Far Eastern Branch, Russian Academy of Science, Vladivostok 690041, Russia}

\author{E.~Pustovalov}
\affiliation{Far Eastern Federal University, 8 Sukhanova Str., 690091 Vladivostok, Russia}

\author{D.~Storozhenko}
\affiliation{Institute of Automation and Control Processes, Far Eastern Branch, Russian Academy of Science, Vladivostok 690041, Russia}

\author{A.~Kuchmizhak}
\email{alex.iacp.dvo@mail.ru}
\affiliation{Institute of Automation and Control Processes, Far Eastern Branch, Russian Academy of Science, Vladivostok 690041, Russia}
\affiliation{Pacific Quantum Center, Far Eastern Federal University, 8 Sukhanova Str., Vladivostok, Russia}

\begin{abstract}
Security labels combining facile structural color readout and physically unclonable one-way function (PUF) approach provide promising strategy for fighting against forgery of marketable products. Here, we justify direct femtosecond-laser printing, a simple and scalable technology, for fabrication of high-resolution (12500 dots per inch) and durable PUF labels with a substantially large encoding capacity of 10$^{895}$ and a simple spectroscopy-free optical signal readout. The proposed tags are comprised of laser-printed plasmonic nanostructures exhibiting unique light scattering behavior and unclonable 3D geometry. Uncontrollable stochastic variation of the nanostructure geometry in the process of their spot-by-spot printing results in random and broadband variation of the scattering color of each laser printed ``pixel'', making laser-printed patterns unique and suitable for PUF labeling.

\end{abstract}

\maketitle
Counterfeiting of various marketable products ranging from luxury items to medical drugs represents global challenge harming patient health and causing significant economic losses~\cite{aldhous2005counterfeit}. Commercialized anti-counterfeiting tags as watermarks, barcodes and holograms were widely applied for labeling of goods owing to their fabrication simplicity and scalability as well as simple authentication procedure~\cite{bansal2013anti}. Meanwhile, security of such tags is basically defined by complexity of the fabrication process as well as uniqueness of the materials used. Owing to these reasons, such tags can be easily duplicated stimulating an active search for novel approaches.

Physical unclonable function (PUF) is based on introduction of stochastic uncontrollable process into the  fabrication of the anti-counterfeit labels making them almost unclonable~\cite{mcgrath2019puf}. Optical PUF labels, i.e. labels with a read-out procedure based on detection of various optical signals as reflection, Rayleigh or Raman scattering, luminescence, etc., have been rapidly evolving during last 
decade~\cite{pappu2002physical,smith2016plasmonic,carro2018optical,hong2020structural,liu2019inkjet,gu2020gap,villegas2021experimental,hu2021flexible}. Utilization of chemically synthesized nanomaterials possessing characteristic fingerprint-like photoluminescence or Raman scattering signals are widely employed to realize optical PUF labels \cite{arppe2017physical,gu2020gap}. Realization of such light-emitting labels is typically technology-free and inexpensive (for example, drop-casting of the nanoparticles over substrates \cite{liu2019inkjet}). However, such labels are formed by chaotic arrangements of nanostructures that significantly complicates algorithms of tag authentication and validation. Moreover, chemically synthesized nanomaterials (e.g. those containing organic compounds) can suffer from photobleaching even under sunlight exposure, while the produced random patterns of nanoparticles can be easily rearranged upon vibration, temperature or liquid impact~\cite{liu2019inkjet}.

\begin{figure}[t!]
\center{\includegraphics[width=0.85\linewidth]{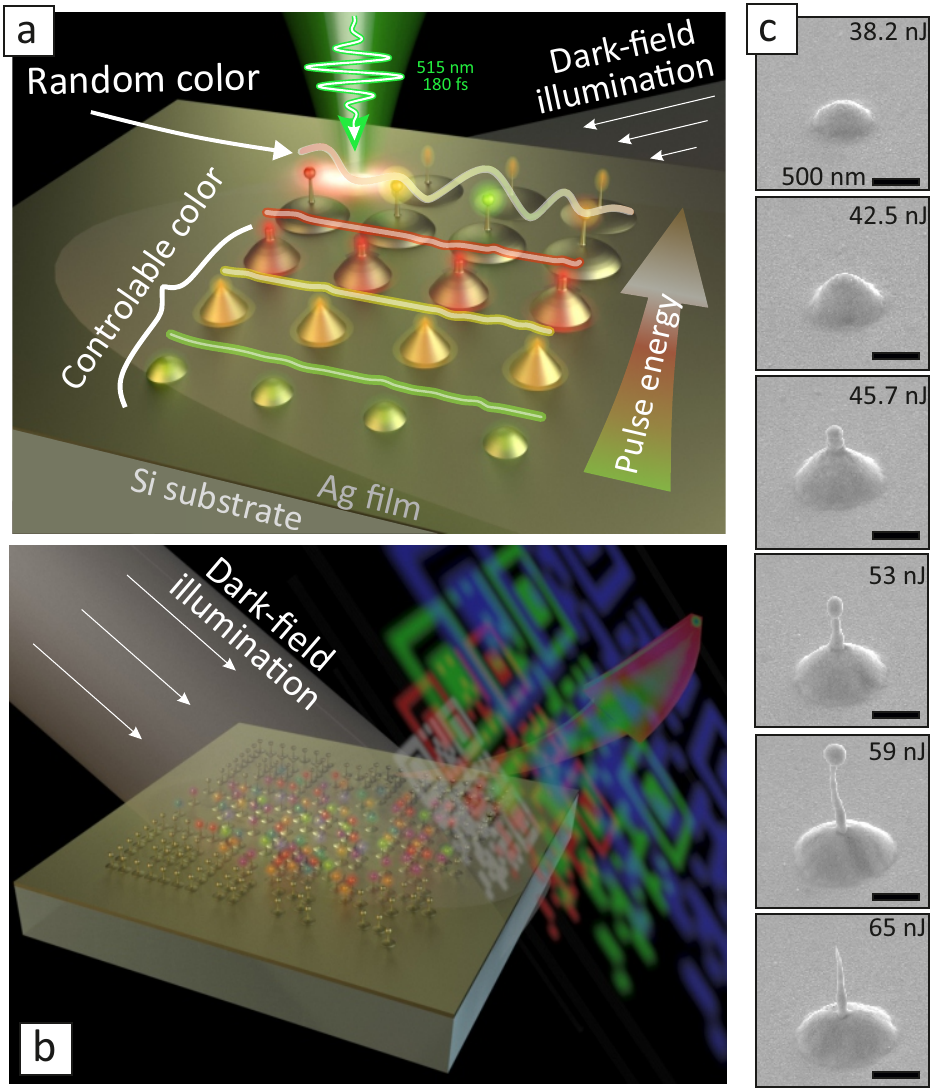}}
\caption{Schematic illustration of (a) fabrication and (b) DF optical read-out of the laser-printed PUF labels. (c) Series of energy-resolved side-view SEM images of the laser-printed Ag nanostructures produced at elevating pulse energy $E$.
}
\label{fig1}
\end{figure}

Nanostructures made of plasmon-active materials or high-index semiconductors possess remarkable optical response owing to resonant nature of their interaction with electromagnetic waves. This interaction coming from resonant absorption/scattering of the optical radiation results in appearance of vivid structural colors that can be identified using common optical microscope in reflection or scattering (dark-field, DF) configurations \cite{kristensen2016plasmonic,xuan2021artificial}. Such structural colors mediated by optically resonant nanostructures can be considered as non-fading and robust against high temperature impact, thus providing ideal platform for anti-counterfeit PUF labeling \cite{hong2020structural}. However, technologies used for fabrication of such nanostructures typically involve multi-step lithography-based procedures that are non-scalable as well as rather expensive, yet provide remarkable control over the geometry of the nanostructures and their optical response. In its turn, excellent reproducibility of the fabrication process renders the related PUF labels easily duplicable even when extremely complicated design of nanostructures is used. In this respect, economically justified technologies allowing to introduce some stochastic process into the PUF labels based on structural colors are highly demanded. Pulsed (femtosecond, fs) laser radiation provides facile and economically justified way for nanostructuring of various materials. Such direct and upscalable processing allows to modify material optical properties at sub-micro length scale opening pathways for structural color information encryption and labeling~\cite{makarov2017light,sakakura2020ultralow}. However, to the best of our knowledge, no attempts were carried out so far to implement direct fs-laser processing for the PUF label realization.

In this Letter, we demonstrate applicability of this method for fabrication of high-resolution PUF labels. Our approach is schematically illustrated in Figure~\ref{fig1}a,b and relies on complex hydrodynamic behaviour of liquid-phase noble-metal thin films (Au, Ag, Cu, etc.) upon their exposure by focused fs laser pulses. In particular, such pulsed laser exposure launches a sequence of rather unique physical processes: (i) local melting of the section of the metal film, (ii) subsequent detachment of the molten shell from the underlying substrate via acoustic relaxation, (iii) accumulation of the molten material at the top of the shell in the form of so-called nanojet, (iv) rearrangement and decay of the nanojet height via Rayleigh-Plateau hydrodynamic instability as well as (v) material resolidification that ''frozens'' the resulting shape of the metal film~\cite{wang2017laser,inogamov2016solitary}. For the fixed size $D_{opt}$ of the laser spot defined by numerical aperture (NA) of the focusing lens, the course of the mentioned processes as well as resulting shape of the imprinted morphology are governed by the applied pulse energy $E$. Generally, this simple and flexibly controlled parameter defines the velocity $\nu$ at which the molten metal shell relaxes from the substrate (silicon, in our case), that governs the resulting geometry of the resolidified surface structure, in its turn. The morphology variation upon corresponding increase of the applied pulse energy $E$ is illustrated by a series of representative side-view SEM images Figure \ref{fig1}c of the nanostructures imprinted on the surface of a 50-nm thick Ag film using 200-fs 515-nm wavelength laser pulses focused at NA=0.3. In general, other noble (or semi-noble) metal films can be used to produce similar structures. However, for demonstration we mainly used Ag owing to its broadband plasmonic response spanning through the entire visible spectral range.

\begin{figure}[t!]
\center{\includegraphics[width=0.95\linewidth]{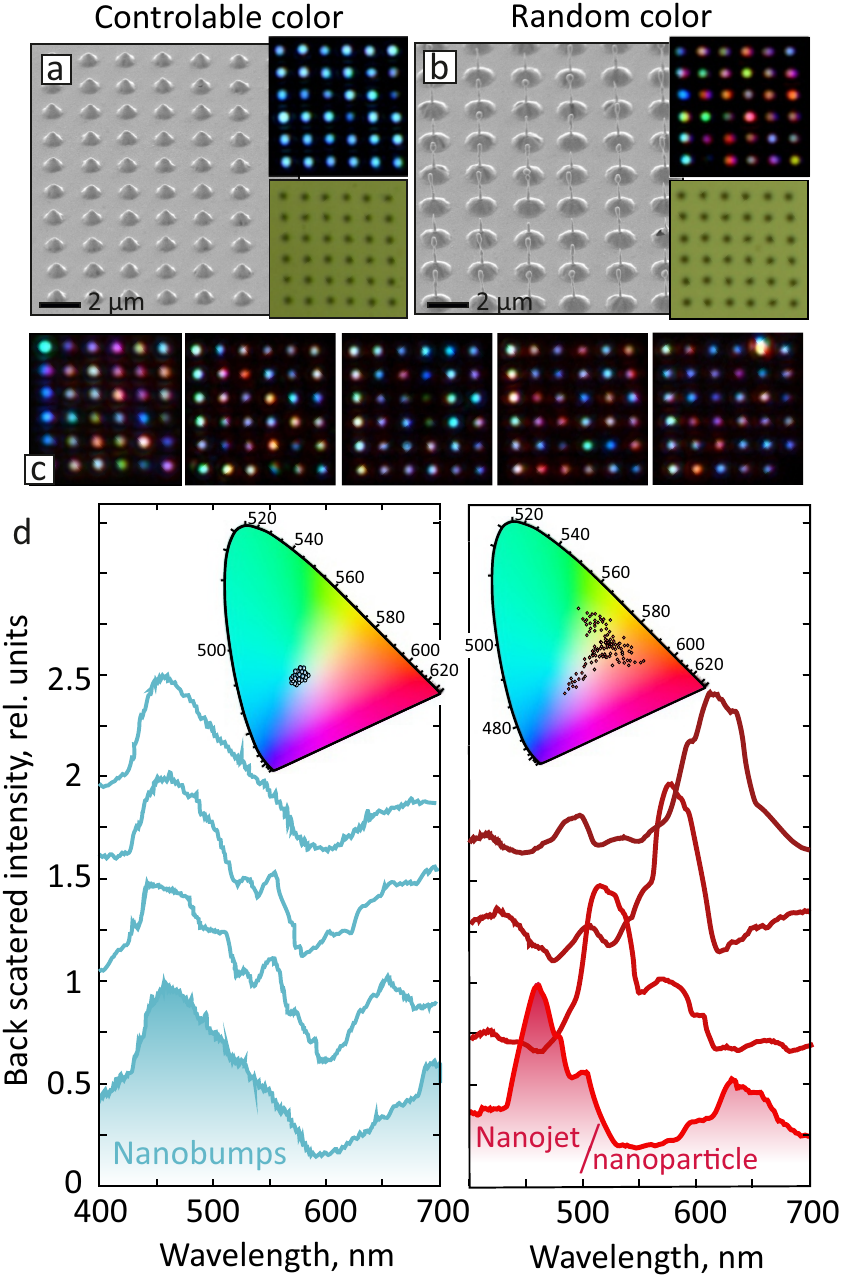}}
\caption{SEM images of square-shape arrays of nanobumps (a) and nanoparticle-nanojet ensembles (b). Insets: DF and BF optical images of the corresponding nanostructure arrays. (c) DF images of the PUF labels (6$\times$6 nanostructures) produced at identical fabrication conditions. (d) DF back-scattered spectra measured from 4 random nanobumps (left; $E$=38.8 nJ) and nanojet-nanoparticle ensembles (right; $E$=57.5 nJ). Insets: representation of the scattering spectra measured from 100 identical nanostructures as coordinates in the CIE 1931 color space.}
\label{fig2}
\end{figure}

First, we highlight an unique 3D morphology of the laser-printed structures that is extremely hard to replicate using any existing lithography-based methods other than mentioned in this paper. Secondly, depending on their shape and the size, produced 3D plasmonic nanostructures resonantly interact with the incident broadband optical radiation that can be observed using common optical microscope with the DF excitation. In particular, when side-illuminated with a white light, the parabola-shape nanobumps (printed at low pulse energies $E$) exhibit bright geometry-dependent scattering originating from excitation of plasmonic modes supported by their hollow shell (Figure \ref{fig2}a). The scattering color gradually red-shifts as the nanobump size increases~\cite{wang2018single}. 3D geometry of the nanostructures reproduces well from one pulse to another resulting in visually identical nanostructures producing identical scattering color in each printed pixel (Figure \ref{fig2}a). However, similar laser-printed nanostructures produced at larger $E$, namely nanojets with a nanoparticle atop (further referred to as nanojet-nanoparticle ensemble), demonstrate completely different scattering behaviour (Figure \ref{fig2}b). In particular, each laser-printed pixel that corresponds to visually identical nanostructure in optical image exhibits bright scattering with its color spanning through the entire visible spectral range defined by plasmonic response of the noble metal used as well as complex multi-resonant structure.

\begin{figure}[t!]
\center{\includegraphics[width=0.99\linewidth]{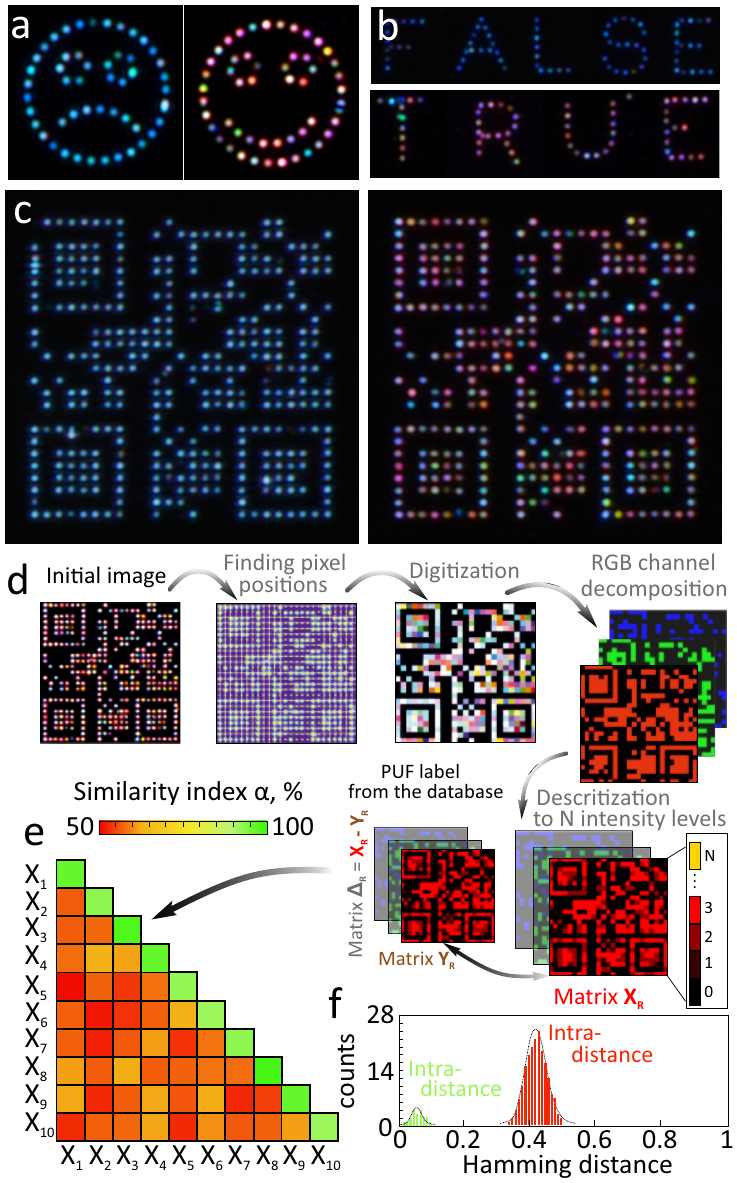}}
\caption{(a-c) DF optical images showing examples of single- and multi-color labels produced with nanobumps and nanojet-nanoparticle ensembles. (d) PUF label authentication and validation algorithm. (e) Distribution of the similarity indexes ($\alpha$) for 10 different PUF labels (QR-codes with 25$\times$25 pixels) calculated based on pairwise comparison. (f) Hamming intra- and inter-distances for 20 different PUF labels (QR-codes, $N$=3).}
\label{fig3}
\end{figure}

Such behavior can be explained by an imperceptible variation of the 3D geometry of the nanojet-nanoparticle ensembles that can not be identified in bright-field (BF) optical microscope Figure \ref{fig2}a,b), yet results in pronounced modification of the scattering color. Such color modification occurs randomly in each pixel owing to simultaneous fluctuation of multiple parameters as local nanoscale roughness of the metal film and morphology modulation of the glass substrate affecting the laser radiation absorption and instrumental short-term pulse-to-pulse stability of the laser system used ($\approx$ 0.05\%). However, cumulative action of all these physical parameters appears to weakly affect morphology of the nanobumps demonstrating almost identical scattering color. The resulting colored pixels can be hardly distinguished from each other, thus can be easily replicated. This feature puts forward the key role of hydrodynamic Rayleigh-Plateau instability, which course is affected much strongly by the other slightly fluctuating physical parameters resulting in a certain random modification of the 3D shape of the the nanojet-nanoparticle ensembles (simultaneous variation of the nanoparticles diameter/shape and the length of the nanojet as well as ejection of the nanoparticle). As a results of these processes, unique physically unclonable 2D pattern of the colored spots can be produced at each fabrication run as it is illustrated by a series of DF optical images. This makes this single-step approach suitable for mass-production of high-resolution PUF tags. The pixel color can be analyzed by measuring the DF scattering spectra from the corresponding plasmonic nanostructures and converting them into the data points of the CIE 1931 color space \cite{wang2018single}. Figure \ref{fig2}(d) compares the representative DF scattering spectra of 10 different nanobumps and nanojet-nanoparticles ensembles printed at fixed $E$=38.8 and 57.5~nJ, respectively. The scattering color of the nanobumps appears to slightly vary from structure to structure resulting in almost identical coordinates in the CIE 1931 color space. In a sharp contrast, color of the nanojet-nanoparticle ensembles printed at fixed $E$ changes remarkably spanning through the entire visible spectral range that results in the significant variation of the color space coordinates (inset in Figure \ref{fig2}d).
As a trade-off between fabrication cost and complexity (Figure \ref{fig3}(a-c)), 2D patterns can be designed on demand tuning the encoding capacity that can be estimated as $L^{n \cdot m}$ ($L$ is number of pixel colors, $m$ is a number of nanostructures per pixel and $n$ is a number of complex pixels). Meanwhile, for practical realization rectangular arrangement of isolated pixels containing the only nanostructure ($m$=1) represents optimal solution allowing easier digitization and faster authentication with minimized level of false positives. Examples of such mono- and random-color PUF labels with rectangular pixel arrangement are shown in Figure \ref{fig3}c. Common authentication approach for such labels includes (i) alignment and determination of each pixel position (including zero-intensity pixels), (ii) digitization of the image, (iii) its subsequent decomposition to the RGB channels as well as (iv) coarse-grained discretization of each pixel intensity (that varied from 0 to 255) to a certain number levels $N$ (see Fig. \ref{fig3}d). To performed the mentioned steps we used Image J and MATLAB R2017 packages.  As a result, each PUF label containing 25$\times$25 pixels was converted into three 2D matrices with a general view

\begin{equation}
X_1^{R,G,B}=\begin{bmatrix}
    x_{1,1}^{R,G,B} & \dots  & x_{1,25}^{R,G,B} \\
        \vdots & \ddots & \vdots \\
    x_{25,1}^{R,G,B} & \dots  & x_{25,25}^{R,G,B}
\end{bmatrix}
\end{equation}

\noindent where x$_{1...25,1...25}^{R,G,B}$ is a certain pixel intensity discretized over $N$ levels for red, green and blue channels. According to the common PUF taxonomy~\cite{mcgrath2019puf}, for our laser-printed labels the Challenge is a pixel position with the input space size of 25$\times$25, while the Response is a coordinate in the RGB color space. In this case, encoding capacity (or total Challenge-Response pair space size) of the PUF label with rectangular pixel arrangement can be assessed as $N^{3n}$  considering independence of 3 RGB channels \cite{gu2020gap}. For example (Figure \ref{fig3}c), 50-$\mu$m$^2$ rectangular-shape QR-code printed at 2-$\mu$m spacing between nanostructures (12,500 DPI resolution unachievable with any drop-casting technique) has the encoding capacity of extraordinary $\approx$10$^{895}$ considering only $N$=3 color levels in each RGB channel that satisfies the modern encoding capacity requirements for anti-counterfeit labels.
Further we used pairwise pixel-by-pixel comparison to assess the difference between 10 random-color PUF labels (QR-codes) printed at fixed fabrication conditions. To do this for each of the RGB channel of two randomly chosen labels characterized by matrices with pixel coordinates y$_{1...25,1...25}^{R,G,B}$ and z$_{1...25,1...25}^{R,G,B}$  we estimated three matrices: 

\noindent
\begin{equation}
\Delta^{R,G,B}=\begin{bmatrix}
    \frac{y_{1,1}^{R,G,B} - z_{1,1}^{R,G,B}}{y_{1,1}^{R,G,B} + z_{1,1}^{R,G,B}}  & \dots  & \frac{y_{1,25}^{R,G,B} - z_{1,25}^{R,G,B}}{y_{1,25}^{R,G,B} + z_{1,25}^{R,G,B}} \\
        \vdots & \ddots & \vdots \\
    \frac{y_{25,1}^{R,G,B} - z_{25,1}^{R,G,B}}{y_{25,1}^{R,G,B} + z_{25,1}^{R,G,B}} & \dots &  \frac{y_{25,25}^{R,G,B} - z_{25,25}^{R,G,B}}{y_{25,25}^{R,G,B} + z_{25,25}^{R,G,B}} .
\end{bmatrix}
\end{equation}

\noindent For the obtained matrices $\Delta^{R,G,B}$, we then calculated averaged similarity index $\alpha$ as a ratio of zero matrix elements to the total number of elements in the matrix. Distribution of similarity indexes is provided in Figure \ref{fig3}e showing that the different PUF labels have the similarity $\approx$ 58$\pm$ 7 \% at $N$=3. The authentication algorithm was further validated by calculating $\alpha$ for a series of 10 images taken for the same PUF QR-codes at slightly varying conditions giving the similarity index of $\approx$ 94$\pm$ 3 \%. Alternatively, for such labels one can calculate a common PUF taxonomy parameter, the Hamming distance, as 1 - $\frac{\alpha}{100\%}$ (Figure \ref{fig3}f). Analysis of the 20 QR-code labels provide the intra- and inter-distances of 0.07$\pm$0.04 and 0.42$\pm$0.08, respectively, demonstrating applicability of the algorithm to reliably distinguish between fake PUF labels and real ones stored in the database. Noteworthy, for illustration purpose we used QR-code design of the PUF labels that contains around one third of the empty (dark) pixels which position is identical for all random realizations of the tested labels. This means that for random-color rectangular-shape nanostructure arrays without empty pixels, the similarity index for different PUF labels will be even lower. In particular, using the similar fabrication approach we produced and analyzed the PUF labels (25$\times$25 nanostructures) without ``dark'' pixels. Such tags still can be easily distinguished using the elaborated algorithm with the estimated inter- and intra-device similarity indexes of 44 and 87\%, respectively. Indeed, lower similarity index and larger encoding capacity can be also achieved by considering larger number of color levels $N$. For example, at $N$=8 and enlarged encoding capacity of 10$^{1693}$, the Hamming intra- and inter-distances become 0.3$\pm$0.1 and 0.6$\pm$0.13 increasing the possibility of false positives/negatives. However, more detailed studies are to be carried out to validate authentication reliability in real-life applications. Along with this, various well-elaborated mathematical methods can be utilized for color matching analysis of optical images \cite{ortiz2019evaluation}, which consideration and optimization fall outside the scope of this Letter.

To conclude, we demonstrated fs-laser printed PUF labels with an large encoding capacity and a simple optical scattering signal readout. Encoding capacity can be increased by increasing the number of random-color pixels $m$. Modern laser technologies provide the multiple methods (DOE-mediated multiplexing or multi-beam interference \cite{pavlov201910,berzins2020direct}) for upscaling the printing rate that allows to increase easily number of random-color pixels $m$ (encoding capacity). If ordering of the pixels is not required, more simple technology-free approach (chaotic speckle-modulated intensity patterns \cite{goodman2007speckle,gurbatov2016lens}) can be utilized. Also, polarization sensitive and form-birefringent laser-printed patterns can recorded to enhance PUF functionality \cite{sakakura2020ultralow}. The average power $P_{av}$ of fs-lasers is following Moore's law of doubling every two years from the beginning of the trend $N = 20$ years ago: $P_{av}\propto 2^{N/2}$. This shows that fs-laser based laser marking with polygon scanners capable of beam travel rates $\sim$km/s is application ready. Noteworthy, the PUF labels produced with noble metal films withstand multiple submerging into the liquids and temperatures up to 400$^\circ$C without any detectable degradation being superior in comparison with any labels utilizing organic probes or nanoparticle drop-cast. The calculated similarity indexes of the images taken from as-fabricated QR-code tags and the same labels stored for 1 year under ambient environment was about 83\% indicating robustness of the PUF labels.
Authors are grateful to Prof. Saulius Juodkazis for valuable discussions and comments. The research is supported by the Russian Science Foundation (21-72-20122).

\section*{Data Availability Statement}
The data that support the findings of this study are available
from the corresponding author upon reasonable request.

\section*{Notes}
There are no conflicts to declare.
%


\begin{thebibliography}{24}%
\makeatletter
\providecommand \@ifxundefined [1]{%
 \@ifx{#1\undefined}
}%
\providecommand \@ifnum [1]{%
 \ifnum #1\expandafter \@firstoftwo
 \else \expandafter \@secondoftwo
 \fi
}%
\providecommand \@ifx [1]{%
 \ifx #1\expandafter \@firstoftwo
 \else \expandafter \@secondoftwo
 \fi
}%
\providecommand \natexlab [1]{#1}%
\providecommand \enquote  [1]{``#1''}%
\providecommand \bibnamefont  [1]{#1}%
\providecommand \bibfnamefont [1]{#1}%
\providecommand \citenamefont [1]{#1}%
\providecommand \href@noop [0]{\@secondoftwo}%
\providecommand \href [0]{\begingroup \@sanitize@url \@href}%
\providecommand \@href[1]{\@@startlink{#1}\@@href}%
\providecommand \@@href[1]{\endgroup#1\@@endlink}%
\providecommand \@sanitize@url [0]{\catcode `\\12\catcode `\$12\catcode
  `\&12\catcode `\#12\catcode `\^12\catcode `\_12\catcode `\%12\relax}%
\providecommand \@@startlink[1]{}%
\providecommand \@@endlink[0]{}%
\providecommand \url  [0]{\begingroup\@sanitize@url \@url }%
\providecommand \@url [1]{\endgroup\@href {#1}{\urlprefix }}%
\providecommand \urlprefix  [0]{URL }%
\providecommand \Eprint [0]{\href }%
\providecommand \doibase [0]{http://dx.doi.org/}%
\providecommand \selectlanguage [0]{\@gobble}%
\providecommand \bibinfo  [0]{\@secondoftwo}%
\providecommand \bibfield  [0]{\@secondoftwo}%
\providecommand \translation [1]{[#1]}%
\providecommand \BibitemOpen [0]{}%
\providecommand \bibitemStop [0]{}%
\providecommand \bibitemNoStop [0]{.\EOS\space}%
\providecommand \EOS [0]{\spacefactor3000\relax}%
\providecommand \BibitemShut  [1]{\csname bibitem#1\endcsname}%
\let\auto@bib@innerbib\@empty
\bibitem [{\citenamefont {Aldhous}(2005)}]{aldhous2005counterfeit}%
  \BibitemOpen
  \bibfield  {author} {\bibinfo {author} {\bibfnamefont {P.}~\bibnamefont
  {Aldhous}},\ }\bibfield  {title} {\enquote {\bibinfo {title} {Counterfeit
  pharmaceuticals: murder by medicine},}\ }\href@noop {} {\bibfield  {journal}
  {\bibinfo  {journal} {Nature}\ }\textbf {\bibinfo {volume} {434}},\ \bibinfo
  {pages} {132--137} (\bibinfo {year} {2005})}\BibitemShut {NoStop}%
\bibitem [{\citenamefont {Bansal}\ \emph {et~al.}(2013)\citenamefont {Bansal},
  \citenamefont {Malla}, \citenamefont {Gudala},\ and\ \citenamefont
  {Tiwari}}]{bansal2013anti}%
  \BibitemOpen
  \bibfield  {author} {\bibinfo {author} {\bibfnamefont {D.}~\bibnamefont
  {Bansal}}, \bibinfo {author} {\bibfnamefont {S.}~\bibnamefont {Malla}},
  \bibinfo {author} {\bibfnamefont {K.}~\bibnamefont {Gudala}}, \ and\ \bibinfo
  {author} {\bibfnamefont {P.}~\bibnamefont {Tiwari}},\ }\bibfield  {title}
  {\enquote {\bibinfo {title} {Anti-counterfeit technologies: a pharmaceutical
  industry perspective},}\ }\href@noop {} {\bibfield  {journal} {\bibinfo
  {journal} {Scientia pharmaceutica}\ }\textbf {\bibinfo {volume} {81}},\
  \bibinfo {pages} {1--14} (\bibinfo {year} {2013})}\BibitemShut {NoStop}%
\bibitem [{\citenamefont {McGrath}\ \emph {et~al.}(2019)\citenamefont
  {McGrath}, \citenamefont {Bagci}, \citenamefont {Wang}, \citenamefont
  {Roedig},\ and\ \citenamefont {Young}}]{mcgrath2019puf}%
  \BibitemOpen
  \bibfield  {author} {\bibinfo {author} {\bibfnamefont {T.}~\bibnamefont
  {McGrath}}, \bibinfo {author} {\bibfnamefont {I.~E.}\ \bibnamefont {Bagci}},
  \bibinfo {author} {\bibfnamefont {Z.~M.}\ \bibnamefont {Wang}}, \bibinfo
  {author} {\bibfnamefont {U.}~\bibnamefont {Roedig}}, \ and\ \bibinfo {author}
  {\bibfnamefont {R.~J.}\ \bibnamefont {Young}},\ }\bibfield  {title} {\enquote
  {\bibinfo {title} {A puf taxonomy},}\ }\href@noop {} {\bibfield  {journal}
  {\bibinfo  {journal} {Applied Physics Reviews}\ }\textbf {\bibinfo {volume}
  {6}},\ \bibinfo {pages} {011303} (\bibinfo {year} {2019})}\BibitemShut
  {NoStop}%
\bibitem [{\citenamefont {Pappu}\ \emph {et~al.}(2002)\citenamefont {Pappu},
  \citenamefont {Recht}, \citenamefont {Taylor},\ and\ \citenamefont
  {Gershenfeld}}]{pappu2002physical}%
  \BibitemOpen
  \bibfield  {author} {\bibinfo {author} {\bibfnamefont {R.}~\bibnamefont
  {Pappu}}, \bibinfo {author} {\bibfnamefont {B.}~\bibnamefont {Recht}},
  \bibinfo {author} {\bibfnamefont {J.}~\bibnamefont {Taylor}}, \ and\ \bibinfo
  {author} {\bibfnamefont {N.}~\bibnamefont {Gershenfeld}},\ }\bibfield
  {title} {\enquote {\bibinfo {title} {Physical one-way functions},}\
  }\href@noop {} {\bibfield  {journal} {\bibinfo  {journal} {Science}\ }\textbf
  {\bibinfo {volume} {297}},\ \bibinfo {pages} {2026--2030} (\bibinfo {year}
  {2002})}\BibitemShut {NoStop}%
\bibitem [{\citenamefont {Smith}, \citenamefont {Patton},\ and\ \citenamefont
  {Skrabalak}(2016)}]{smith2016plasmonic}%
  \BibitemOpen
  \bibfield  {author} {\bibinfo {author} {\bibfnamefont {A.~F.}\ \bibnamefont
  {Smith}}, \bibinfo {author} {\bibfnamefont {P.}~\bibnamefont {Patton}}, \
  and\ \bibinfo {author} {\bibfnamefont {S.~E.}\ \bibnamefont {Skrabalak}},\
  }\bibfield  {title} {\enquote {\bibinfo {title} {Plasmonic nanoparticles as a
  physically unclonable function for responsive anti-counterfeit
  nanofingerprints},}\ }\href@noop {} {\bibfield  {journal} {\bibinfo
  {journal} {Advanced Functional Materials}\ }\textbf {\bibinfo {volume}
  {26}},\ \bibinfo {pages} {1315--1321} (\bibinfo {year} {2016})}\BibitemShut
  {NoStop}%
\bibitem [{\citenamefont {Carro-Temboury}\ \emph {et~al.}(2018)\citenamefont
  {Carro-Temboury}, \citenamefont {Arppe}, \citenamefont {Vosch},\ and\
  \citenamefont {S{\o}rensen}}]{carro2018optical}%
  \BibitemOpen
  \bibfield  {author} {\bibinfo {author} {\bibfnamefont {M.~R.}\ \bibnamefont
  {Carro-Temboury}}, \bibinfo {author} {\bibfnamefont {R.}~\bibnamefont
  {Arppe}}, \bibinfo {author} {\bibfnamefont {T.}~\bibnamefont {Vosch}}, \ and\
  \bibinfo {author} {\bibfnamefont {T.~J.}\ \bibnamefont {S{\o}rensen}},\
  }\bibfield  {title} {\enquote {\bibinfo {title} {An optical authentication
  system based on imaging of excitation-selected lanthanide luminescence},}\
  }\href@noop {} {\bibfield  {journal} {\bibinfo  {journal} {Science advances}\
  }\textbf {\bibinfo {volume} {4}},\ \bibinfo {pages} {e1701384} (\bibinfo
  {year} {2018})}\BibitemShut {NoStop}%
\bibitem [{\citenamefont {Hong}, \citenamefont {Yuan},\ and\ \citenamefont
  {Chen}(2020)}]{hong2020structural}%
  \BibitemOpen
  \bibfield  {author} {\bibinfo {author} {\bibfnamefont {W.}~\bibnamefont
  {Hong}}, \bibinfo {author} {\bibfnamefont {Z.}~\bibnamefont {Yuan}}, \ and\
  \bibinfo {author} {\bibfnamefont {X.}~\bibnamefont {Chen}},\ }\bibfield
  {title} {\enquote {\bibinfo {title} {Structural color materials for optical
  anticounterfeiting},}\ }\href@noop {} {\bibfield  {journal} {\bibinfo
  {journal} {Small}\ }\textbf {\bibinfo {volume} {16}},\ \bibinfo {pages}
  {1907626} (\bibinfo {year} {2020})}\BibitemShut {NoStop}%
\bibitem [{\citenamefont {Liu}\ \emph {et~al.}(2019)\citenamefont {Liu},
  \citenamefont {Han}, \citenamefont {Li}, \citenamefont {Zhao}, \citenamefont
  {Chen}, \citenamefont {Xu}, \citenamefont {Zheng}, \citenamefont {Hu},
  \citenamefont {Yao}, \citenamefont {Guo} \emph {et~al.}}]{liu2019inkjet}%
  \BibitemOpen
  \bibfield  {author} {\bibinfo {author} {\bibfnamefont {Y.}~\bibnamefont
  {Liu}}, \bibinfo {author} {\bibfnamefont {F.}~\bibnamefont {Han}}, \bibinfo
  {author} {\bibfnamefont {F.}~\bibnamefont {Li}}, \bibinfo {author}
  {\bibfnamefont {Y.}~\bibnamefont {Zhao}}, \bibinfo {author} {\bibfnamefont
  {M.}~\bibnamefont {Chen}}, \bibinfo {author} {\bibfnamefont {Z.}~\bibnamefont
  {Xu}}, \bibinfo {author} {\bibfnamefont {X.}~\bibnamefont {Zheng}}, \bibinfo
  {author} {\bibfnamefont {H.}~\bibnamefont {Hu}}, \bibinfo {author}
  {\bibfnamefont {J.}~\bibnamefont {Yao}}, \bibinfo {author} {\bibfnamefont
  {T.}~\bibnamefont {Guo}},  \emph {et~al.},\ }\bibfield  {title} {\enquote
  {\bibinfo {title} {Inkjet-printed unclonable quantum dot fluorescent
  anti-counterfeiting labels with artificial intelligence authentication},}\
  }\href@noop {} {\bibfield  {journal} {\bibinfo  {journal} {Nature
  communications}\ }\textbf {\bibinfo {volume} {10}},\ \bibinfo {pages} {1--9}
  (\bibinfo {year} {2019})}\BibitemShut {NoStop}%
\bibitem [{\citenamefont {Gu}\ \emph {et~al.}(2020)\citenamefont {Gu},
  \citenamefont {He}, \citenamefont {Zhang}, \citenamefont {Lin}, \citenamefont
  {Thackray},\ and\ \citenamefont {Ye}}]{gu2020gap}%
  \BibitemOpen
  \bibfield  {author} {\bibinfo {author} {\bibfnamefont {Y.}~\bibnamefont
  {Gu}}, \bibinfo {author} {\bibfnamefont {C.}~\bibnamefont {He}}, \bibinfo
  {author} {\bibfnamefont {Y.}~\bibnamefont {Zhang}}, \bibinfo {author}
  {\bibfnamefont {L.}~\bibnamefont {Lin}}, \bibinfo {author} {\bibfnamefont
  {B.~D.}\ \bibnamefont {Thackray}}, \ and\ \bibinfo {author} {\bibfnamefont
  {J.}~\bibnamefont {Ye}},\ }\bibfield  {title} {\enquote {\bibinfo {title}
  {Gap-enhanced raman tags for physically unclonable anticounterfeiting
  labels},}\ }\href@noop {} {\bibfield  {journal} {\bibinfo  {journal} {Nature
  communications}\ }\textbf {\bibinfo {volume} {11}},\ \bibinfo {pages} {1--13}
  (\bibinfo {year} {2020})}\BibitemShut {NoStop}%
\bibitem [{\citenamefont {Villegas}, \citenamefont {Paredes},\ and\
  \citenamefont {Rasras}(2021)}]{villegas2021experimental}%
  \BibitemOpen
  \bibfield  {author} {\bibinfo {author} {\bibfnamefont {J.~E.}\ \bibnamefont
  {Villegas}}, \bibinfo {author} {\bibfnamefont {B.}~\bibnamefont {Paredes}}, \
  and\ \bibinfo {author} {\bibfnamefont {M.}~\bibnamefont {Rasras}},\
  }\bibfield  {title} {\enquote {\bibinfo {title} {Experimental studies of
  plasmonics-enhanced optical physically unclonable functions},}\ }\href@noop
  {} {\bibfield  {journal} {\bibinfo  {journal} {Optics Express}\ }\textbf
  {\bibinfo {volume} {29}},\ \bibinfo {pages} {32020--32030} (\bibinfo {year}
  {2021})}\BibitemShut {NoStop}%
\bibitem [{\citenamefont {Hu}\ \emph {et~al.}(2021)\citenamefont {Hu},
  \citenamefont {Zhang}, \citenamefont {Wang}, \citenamefont {Liu},
  \citenamefont {Sun}, \citenamefont {Li}, \citenamefont {Lv}, \citenamefont
  {Liang}, \citenamefont {Jiao}, \citenamefont {Zhao} \emph
  {et~al.}}]{hu2021flexible}%
  \BibitemOpen
  \bibfield  {author} {\bibinfo {author} {\bibfnamefont {Y.-W.}\ \bibnamefont
  {Hu}}, \bibinfo {author} {\bibfnamefont {T.-P.}\ \bibnamefont {Zhang}},
  \bibinfo {author} {\bibfnamefont {C.-F.}\ \bibnamefont {Wang}}, \bibinfo
  {author} {\bibfnamefont {K.-K.}\ \bibnamefont {Liu}}, \bibinfo {author}
  {\bibfnamefont {Y.}~\bibnamefont {Sun}}, \bibinfo {author} {\bibfnamefont
  {L.}~\bibnamefont {Li}}, \bibinfo {author} {\bibfnamefont {C.-F.}\
  \bibnamefont {Lv}}, \bibinfo {author} {\bibfnamefont {Y.-C.}\ \bibnamefont
  {Liang}}, \bibinfo {author} {\bibfnamefont {F.-H.}\ \bibnamefont {Jiao}},
  \bibinfo {author} {\bibfnamefont {W.-B.}\ \bibnamefont {Zhao}},  \emph
  {et~al.},\ }\bibfield  {title} {\enquote {\bibinfo {title} {Flexible and
  biocompatible physical unclonable function anti-counterfeiting label},}\
  }\href@noop {} {\bibfield  {journal} {\bibinfo  {journal} {Advanced
  Functional Materials}\ ,\ \bibinfo {pages} {2102108}} (\bibinfo {year}
  {2021})}\BibitemShut {NoStop}%
\bibitem [{\citenamefont {Arppe}\ and\ \citenamefont
  {S{\o}rensen}(2017)}]{arppe2017physical}%
  \BibitemOpen
  \bibfield  {author} {\bibinfo {author} {\bibfnamefont {R.}~\bibnamefont
  {Arppe}}\ and\ \bibinfo {author} {\bibfnamefont {T.~J.}\ \bibnamefont
  {S{\o}rensen}},\ }\bibfield  {title} {\enquote {\bibinfo {title} {Physical
  unclonable functions generated through chemical methods for
  anti-counterfeiting},}\ }\href@noop {} {\bibfield  {journal} {\bibinfo
  {journal} {Nature Reviews Chemistry}\ }\textbf {\bibinfo {volume} {1}},\
  \bibinfo {pages} {1--13} (\bibinfo {year} {2017})}\BibitemShut {NoStop}%
\bibitem [{\citenamefont {Kristensen}\ \emph {et~al.}(2016)\citenamefont
  {Kristensen}, \citenamefont {Yang}, \citenamefont {Bozhevolnyi},
  \citenamefont {Link}, \citenamefont {Nordlander}, \citenamefont {Halas},\
  and\ \citenamefont {Mortensen}}]{kristensen2016plasmonic}%
  \BibitemOpen
  \bibfield  {author} {\bibinfo {author} {\bibfnamefont {A.}~\bibnamefont
  {Kristensen}}, \bibinfo {author} {\bibfnamefont {J.~K.}\ \bibnamefont
  {Yang}}, \bibinfo {author} {\bibfnamefont {S.~I.}\ \bibnamefont
  {Bozhevolnyi}}, \bibinfo {author} {\bibfnamefont {S.}~\bibnamefont {Link}},
  \bibinfo {author} {\bibfnamefont {P.}~\bibnamefont {Nordlander}}, \bibinfo
  {author} {\bibfnamefont {N.~J.}\ \bibnamefont {Halas}}, \ and\ \bibinfo
  {author} {\bibfnamefont {N.~A.}\ \bibnamefont {Mortensen}},\ }\bibfield
  {title} {\enquote {\bibinfo {title} {Plasmonic colour generation},}\
  }\href@noop {} {\bibfield  {journal} {\bibinfo  {journal} {Nature Reviews
  Materials}\ }\textbf {\bibinfo {volume} {2}},\ \bibinfo {pages} {1--14}
  (\bibinfo {year} {2016})}\BibitemShut {NoStop}%
\bibitem [{\citenamefont {Xuan}\ \emph {et~al.}(2021)\citenamefont {Xuan},
  \citenamefont {Li}, \citenamefont {Liu}, \citenamefont {Yi}, \citenamefont
  {Wang},\ and\ \citenamefont {Lu}}]{xuan2021artificial}%
  \BibitemOpen
  \bibfield  {author} {\bibinfo {author} {\bibfnamefont {Z.}~\bibnamefont
  {Xuan}}, \bibinfo {author} {\bibfnamefont {J.}~\bibnamefont {Li}}, \bibinfo
  {author} {\bibfnamefont {Q.}~\bibnamefont {Liu}}, \bibinfo {author}
  {\bibfnamefont {F.}~\bibnamefont {Yi}}, \bibinfo {author} {\bibfnamefont
  {S.}~\bibnamefont {Wang}}, \ and\ \bibinfo {author} {\bibfnamefont
  {W.}~\bibnamefont {Lu}},\ }\bibfield  {title} {\enquote {\bibinfo {title}
  {Artificial structural colors and applications},}\ }\href@noop {} {\bibfield
  {journal} {\bibinfo  {journal} {The Innovation}\ ,\ \bibinfo {pages}
  {100081}} (\bibinfo {year} {2021})}\BibitemShut {NoStop}%
\bibitem [{\citenamefont {Makarov}\ \emph {et~al.}(2017)\citenamefont
  {Makarov}, \citenamefont {Zalogina}, \citenamefont {Tajik}, \citenamefont
  {Zuev}, \citenamefont {Rybin}, \citenamefont {Kuchmizhak}, \citenamefont
  {Juodkazis},\ and\ \citenamefont {Kivshar}}]{makarov2017light}%
  \BibitemOpen
  \bibfield  {author} {\bibinfo {author} {\bibfnamefont {S.~V.}\ \bibnamefont
  {Makarov}}, \bibinfo {author} {\bibfnamefont {A.~S.}\ \bibnamefont
  {Zalogina}}, \bibinfo {author} {\bibfnamefont {M.}~\bibnamefont {Tajik}},
  \bibinfo {author} {\bibfnamefont {D.~A.}\ \bibnamefont {Zuev}}, \bibinfo
  {author} {\bibfnamefont {M.~V.}\ \bibnamefont {Rybin}}, \bibinfo {author}
  {\bibfnamefont {A.~A.}\ \bibnamefont {Kuchmizhak}}, \bibinfo {author}
  {\bibfnamefont {S.}~\bibnamefont {Juodkazis}}, \ and\ \bibinfo {author}
  {\bibfnamefont {Y.}~\bibnamefont {Kivshar}},\ }\bibfield  {title} {\enquote
  {\bibinfo {title} {Light-induced tuning and reconfiguration of nanophotonic
  structures},}\ }\href@noop {} {\bibfield  {journal} {\bibinfo  {journal}
  {Laser \& Photonics Reviews}\ }\textbf {\bibinfo {volume} {11}},\ \bibinfo
  {pages} {1700108} (\bibinfo {year} {2017})}\BibitemShut {NoStop}%
\bibitem [{\citenamefont {Sakakura}\ \emph {et~al.}(2020)\citenamefont
  {Sakakura}, \citenamefont {Lei}, \citenamefont {Wang}, \citenamefont {Yu},\
  and\ \citenamefont {Kazansky}}]{sakakura2020ultralow}%
  \BibitemOpen
  \bibfield  {author} {\bibinfo {author} {\bibfnamefont {M.}~\bibnamefont
  {Sakakura}}, \bibinfo {author} {\bibfnamefont {Y.}~\bibnamefont {Lei}},
  \bibinfo {author} {\bibfnamefont {L.}~\bibnamefont {Wang}}, \bibinfo {author}
  {\bibfnamefont {Y.-H.}\ \bibnamefont {Yu}}, \ and\ \bibinfo {author}
  {\bibfnamefont {P.~G.}\ \bibnamefont {Kazansky}},\ }\bibfield  {title}
  {\enquote {\bibinfo {title} {Ultralow-loss geometric phase and polarization
  shaping by ultrafast laser writing in silica glass},}\ }\href@noop {}
  {\bibfield  {journal} {\bibinfo  {journal} {Light: Science \& Applications}\
  }\textbf {\bibinfo {volume} {9}},\ \bibinfo {pages} {1--10} (\bibinfo {year}
  {2020})}\BibitemShut {NoStop}%
\bibitem [{\citenamefont {Wang}\ \emph {et~al.}(2017)\citenamefont {Wang},
  \citenamefont {Kuchmizhak}, \citenamefont {Li}, \citenamefont {Juodkazis},
  \citenamefont {Vitrik}, \citenamefont {Kulchin}, \citenamefont {Zhakhovsky},
  \citenamefont {Danilov}, \citenamefont {Ionin}, \citenamefont {Kudryashov}
  \emph {et~al.}}]{wang2017laser}%
  \BibitemOpen
  \bibfield  {author} {\bibinfo {author} {\bibfnamefont {X.}~\bibnamefont
  {Wang}}, \bibinfo {author} {\bibfnamefont {A.}~\bibnamefont {Kuchmizhak}},
  \bibinfo {author} {\bibfnamefont {X.}~\bibnamefont {Li}}, \bibinfo {author}
  {\bibfnamefont {S.}~\bibnamefont {Juodkazis}}, \bibinfo {author}
  {\bibfnamefont {O.}~\bibnamefont {Vitrik}}, \bibinfo {author} {\bibfnamefont
  {Y.~N.}\ \bibnamefont {Kulchin}}, \bibinfo {author} {\bibfnamefont
  {V.}~\bibnamefont {Zhakhovsky}}, \bibinfo {author} {\bibfnamefont
  {P.}~\bibnamefont {Danilov}}, \bibinfo {author} {\bibfnamefont
  {A.}~\bibnamefont {Ionin}}, \bibinfo {author} {\bibfnamefont
  {S.}~\bibnamefont {Kudryashov}},  \emph {et~al.},\ }\bibfield  {title}
  {\enquote {\bibinfo {title} {Laser-induced translative hydrodynamic mass
  snapshots: noninvasive characterization and predictive modeling via mapping
  at nanoscale},}\ }\href@noop {} {\bibfield  {journal} {\bibinfo  {journal}
  {Physical Review Applied}\ }\textbf {\bibinfo {volume} {8}},\ \bibinfo
  {pages} {044016} (\bibinfo {year} {2017})}\BibitemShut {NoStop}%
\bibitem [{\citenamefont {Inogamov}\ \emph {et~al.}(2016)\citenamefont
  {Inogamov}, \citenamefont {Zhakhovsky}, \citenamefont {Khokhlov},
  \citenamefont {Petrov},\ and\ \citenamefont {Migdal}}]{inogamov2016solitary}%
  \BibitemOpen
  \bibfield  {author} {\bibinfo {author} {\bibfnamefont {N.~A.}\ \bibnamefont
  {Inogamov}}, \bibinfo {author} {\bibfnamefont {V.~V.}\ \bibnamefont
  {Zhakhovsky}}, \bibinfo {author} {\bibfnamefont {V.~A.}\ \bibnamefont
  {Khokhlov}}, \bibinfo {author} {\bibfnamefont {Y.~V.}\ \bibnamefont
  {Petrov}}, \ and\ \bibinfo {author} {\bibfnamefont {K.~P.}\ \bibnamefont
  {Migdal}},\ }\bibfield  {title} {\enquote {\bibinfo {title} {Solitary
  nanostructures produced by ultrashort laser pulse},}\ }\href@noop {}
  {\bibfield  {journal} {\bibinfo  {journal} {Nanoscale research letters}\
  }\textbf {\bibinfo {volume} {11}},\ \bibinfo {pages} {1--13} (\bibinfo {year}
  {2016})}\BibitemShut {NoStop}%
\bibitem [{\citenamefont {Wang}\ \emph {et~al.}(2018)\citenamefont {Wang},
  \citenamefont {Kuchmizhak}, \citenamefont {Storozhenko}, \citenamefont
  {Makarov},\ and\ \citenamefont {Juodkazis}}]{wang2018single}%
  \BibitemOpen
  \bibfield  {author} {\bibinfo {author} {\bibfnamefont {X.}~\bibnamefont
  {Wang}}, \bibinfo {author} {\bibfnamefont {A.}~\bibnamefont {Kuchmizhak}},
  \bibinfo {author} {\bibfnamefont {D.}~\bibnamefont {Storozhenko}}, \bibinfo
  {author} {\bibfnamefont {S.}~\bibnamefont {Makarov}}, \ and\ \bibinfo
  {author} {\bibfnamefont {S.}~\bibnamefont {Juodkazis}},\ }\bibfield  {title}
  {\enquote {\bibinfo {title} {Single-step laser plasmonic coloration of metal
  films},}\ }\href@noop {} {\bibfield  {journal} {\bibinfo  {journal} {ACS
  applied materials \& interfaces}\ }\textbf {\bibinfo {volume} {10}},\
  \bibinfo {pages} {1422--1427} (\bibinfo {year} {2018})}\BibitemShut {NoStop}%
\bibitem [{\citenamefont {Ortiz-Jaramillo}\ \emph {et~al.}(2019)\citenamefont
  {Ortiz-Jaramillo}, \citenamefont {Kumcu}, \citenamefont {Platisa},\ and\
  \citenamefont {Philips}}]{ortiz2019evaluation}%
  \BibitemOpen
  \bibfield  {author} {\bibinfo {author} {\bibfnamefont {B.}~\bibnamefont
  {Ortiz-Jaramillo}}, \bibinfo {author} {\bibfnamefont {A.}~\bibnamefont
  {Kumcu}}, \bibinfo {author} {\bibfnamefont {L.}~\bibnamefont {Platisa}}, \
  and\ \bibinfo {author} {\bibfnamefont {W.}~\bibnamefont {Philips}},\
  }\bibfield  {title} {\enquote {\bibinfo {title} {Evaluation of color
  differences in natural scene color images},}\ }\href@noop {} {\bibfield
  {journal} {\bibinfo  {journal} {Signal Processing: Image Communication}\
  }\textbf {\bibinfo {volume} {71}},\ \bibinfo {pages} {128--137} (\bibinfo
  {year} {2019})}\BibitemShut {NoStop}%
\bibitem [{\citenamefont {Pavlov}\ \emph {et~al.}(2019)\citenamefont {Pavlov},
  \citenamefont {Gurbatov}, \citenamefont {Kudryashov}, \citenamefont
  {Danilov}, \citenamefont {Porfirev}, \citenamefont {Khonina}, \citenamefont
  {Vitrik}, \citenamefont {Kulinich}, \citenamefont {Lapine},\ and\
  \citenamefont {Kuchmizhak}}]{pavlov201910}%
  \BibitemOpen
  \bibfield  {author} {\bibinfo {author} {\bibfnamefont {D.}~\bibnamefont
  {Pavlov}}, \bibinfo {author} {\bibfnamefont {S.}~\bibnamefont {Gurbatov}},
  \bibinfo {author} {\bibfnamefont {S.}~\bibnamefont {Kudryashov}}, \bibinfo
  {author} {\bibfnamefont {P.}~\bibnamefont {Danilov}}, \bibinfo {author}
  {\bibfnamefont {A.}~\bibnamefont {Porfirev}}, \bibinfo {author}
  {\bibfnamefont {S.}~\bibnamefont {Khonina}}, \bibinfo {author} {\bibfnamefont
  {O.}~\bibnamefont {Vitrik}}, \bibinfo {author} {\bibfnamefont
  {S.}~\bibnamefont {Kulinich}}, \bibinfo {author} {\bibfnamefont
  {M.}~\bibnamefont {Lapine}}, \ and\ \bibinfo {author} {\bibfnamefont
  {A.}~\bibnamefont {Kuchmizhak}},\ }\bibfield  {title} {\enquote {\bibinfo
  {title} {10-million-elements-per-second printing of infrared-resonant
  plasmonic arrays by multiplexed laser pulses},}\ }\href@noop {} {\bibfield
  {journal} {\bibinfo  {journal} {Optics letters}\ }\textbf {\bibinfo {volume}
  {44}},\ \bibinfo {pages} {283--286} (\bibinfo {year} {2019})}\BibitemShut
  {NoStop}%
\bibitem [{\citenamefont {Berzins}\ \emph {et~al.}(2020)\citenamefont
  {Berzins}, \citenamefont {Indrisiunas}, \citenamefont {Van~Erve},
  \citenamefont {Nagarajan}, \citenamefont {Fasold}, \citenamefont {Steinert},
  \citenamefont {Gerini}, \citenamefont {Gecys}, \citenamefont {Pertsch},
  \citenamefont {Ba?umer} \emph {et~al.}}]{berzins2020direct}%
  \BibitemOpen
  \bibfield  {author} {\bibinfo {author} {\bibfnamefont {J.}~\bibnamefont
  {Berzins}}, \bibinfo {author} {\bibfnamefont {S.}~\bibnamefont
  {Indrisiunas}}, \bibinfo {author} {\bibfnamefont {K.}~\bibnamefont
  {Van~Erve}}, \bibinfo {author} {\bibfnamefont {A.}~\bibnamefont {Nagarajan}},
  \bibinfo {author} {\bibfnamefont {S.}~\bibnamefont {Fasold}}, \bibinfo
  {author} {\bibfnamefont {M.}~\bibnamefont {Steinert}}, \bibinfo {author}
  {\bibfnamefont {G.}~\bibnamefont {Gerini}}, \bibinfo {author} {\bibfnamefont
  {P.}~\bibnamefont {Gecys}}, \bibinfo {author} {\bibfnamefont
  {T.}~\bibnamefont {Pertsch}}, \bibinfo {author} {\bibfnamefont {S.~M.}\
  \bibnamefont {Ba?umer}},  \emph {et~al.},\ }\bibfield  {title} {\enquote
  {\bibinfo {title} {Direct and high-throughput fabrication of mie-resonant
  metasurfaces via single-pulse laser interference},}\ }\href@noop {}
  {\bibfield  {journal} {\bibinfo  {journal} {ACS nano}\ }\textbf {\bibinfo
  {volume} {14}},\ \bibinfo {pages} {6138--6149} (\bibinfo {year}
  {2020})}\BibitemShut {NoStop}%
\bibitem [{\citenamefont {Goodman}(2007)}]{goodman2007speckle}%
  \BibitemOpen
  \bibfield  {author} {\bibinfo {author} {\bibfnamefont {J.~W.}\ \bibnamefont
  {Goodman}},\ }\href@noop {} {\emph {\bibinfo {title} {Speckle phenomena in
  optics: theory and applications}}}\ (\bibinfo  {publisher} {Roberts and
  Company Publishers},\ \bibinfo {year} {2007})\BibitemShut {NoStop}%
\bibitem [{\citenamefont {Gurbatov}\ \emph {et~al.}(2016)\citenamefont
  {Gurbatov}, \citenamefont {Kuchmizhak}, \citenamefont {Vitrik},\ and\
  \citenamefont {Kulchin}}]{gurbatov2016lens}%
  \BibitemOpen
  \bibfield  {author} {\bibinfo {author} {\bibfnamefont {S.}~\bibnamefont
  {Gurbatov}}, \bibinfo {author} {\bibfnamefont {A.}~\bibnamefont
  {Kuchmizhak}}, \bibinfo {author} {\bibfnamefont {O.}~\bibnamefont {Vitrik}},
  \ and\ \bibinfo {author} {\bibfnamefont {Y.}~\bibnamefont {Kulchin}},\
  }\bibfield  {title} {\enquote {\bibinfo {title} {Lens-free laser
  nanopatterning of large-scale metal film areas with structured light for
  biosensing applications},}\ }\href@noop {} {\bibfield  {journal} {\bibinfo
  {journal} {Optics express}\ }\textbf {\bibinfo {volume} {24}},\ \bibinfo
  {pages} {18898--18906} (\bibinfo {year} {2016})}\BibitemShut {NoStop}%
\end{thebibliography}
 \end{document}